\documentclass[twocolumn]{aastex631}
\usepackage[version=4]{mhchem}
\usepackage{subfigure}
\usepackage{hyperref}
\usepackage{natbib}
\usepackage{booktabs} 
\usepackage{xcolor}

\shorttitle{Methylated Biosignatures on Hycean Worlds}
\shortauthors{Leung et al.}

\graphicspath{{./}{figures/}}

\usepackage{mhchem}

\begin{document}

\title{Examining the Potential for Methyl Halide Accumulation and Detectability in Possible Hycean-Type Atmospheres}

\correspondingauthor{Michaela Leung}
\email{michaela.leung@email.ucr.edu}

\author[0000-0003-1906-5093]{Michaela Leung}
\affiliation{Department of Earth and Planetary Sciences, University of California, Riverside, California, 92521}
\affiliation{NASA Alternative Earths Team}

\author{Shang-Min Tsai}
\affiliation{Department of Earth and Planetary Sciences, University of California, Riverside, California, 92521}
\affiliation{NASA Alternative Earths Team}

\author[0000-0002-2949-2163]{Edward W. Schwieterman}
\affiliation{Department of Earth and Planetary Sciences, University of California, Riverside, California, 92521}
\affiliation{Blue Marble Space Institute of Science, Seattle, WA, USA}

\author{Daniel Angerhausen}
\affiliation{Blue Marble Space Institute of Science, Seattle, WA, USA}
\affiliation{ETH Zurich, Institute for Particle Physics \& Astrophysics, Wolfgang-Pauli-Str. 27, 8093 Zurich, Switzerland}

\author{Janina Hansen}
\affiliation{ETH Zurich, Institute for Particle Physics \& Astrophysics, Wolfgang-Pauli-Str. 27, 8093 Zurich, Switzerland}


\begin{abstract}

Some sub-Neptune planets may host habitable conditions; for example "Hycean" worlds with \ce{H2} envelopes over liquid water oceans can maintain potentially hospitable pressures and temperatures at their surface. Recent JWST observations of K2-18b and TOI-270d have shown that such worlds could be compelling targets for biosignature searches, given their extended scale heights and therefore large atmospheric signatures. Methylated biosignatures, a broad group of gases that can be generated by biological attachment of a \ce{CH3} group to an environmental substrate, have been proposed as candidate signs of life for Earth-like exoplanets. However, methyl halides (\ce{CH3} + halogen) have not yet been robustly examined with self-consistent photochemical and spectral models for planets with \ce{H2}-dominated atmospheres. Here we demonstrate that methyl chloride (\ce{CH3Cl}), predominantly produced by marine microbes, could be detected using JWST in tens of transits or fewer for Hycean planets, comparable to detection requirements for other potential atmospheric biosignatures. The threshold atmospheric mixing ratio for detectability is $\sim$10 ppm, which can accumulate with global fluxes comparable to moderately productive local environments on Earth. 
\end{abstract}



\section{Introduction} \label{sec:intro} 
The search for life beyond the Earth is a compelling motivation to develop observational and modeling tools to characterize sub-Neptune and smaller exoplanets with high fidelity. Instrumental constraints have until recently limited this speculative area to preparatory modeling work, influenced by the planetary mass, radius, and distribution information gathered by survey missions such as Kepler \citep[i.e.,][]{Greiss2012-bx, Greiss2012-dj} and TESS \citep[i.e.,][]{Sharma2017-mm, Guerrero2021-ox}. These missions have revolutionized our statistical understanding of the exoplanet population, revealing the high frequency of ``sub-Neptune" planets that exist between the radii of terrestrial Earth-like and gaseous Neptune-like planets \citep[i.e.,][]{Rivera2005, Valencia2007-sg, Bergsten2022-vw}. Substantial theoretical attention has been applied to these planets and their conditions, including their potential habitability \citep[i.e.,][]{Von_Bloh2009-jh, Hu2019-qu, Claudi2020-xt}. A well-explored conception of these planets as conventionally habitable invokes an \ce{H2} rich atmosphere above a surface liquid water ocean , accommodating the reported mass and radius measurements, and supporting a potentially hospitable ocean \citep{Madhusudhan2021-io}. 

While some temperate sub-Neptune planets may support habitable (Hycean) conditions, this prediction is still being tested against observational data. Preliminary observations with JWST are degenerate with multiple atmospheric models that fit the limited data, including those that preclude habitable environments \citep{Biagini2024-ru, Damiano2024-wo}. It is difficult to differentiate between plausible scenarios suggested by interior and atmospheric models \citep[i.e.,][]{Madhusudhan2020-kn}, especially without a corresponding analog in our solar system. One method of distinguishing a massive envelope from a shallow habitable condition relies on differential solubility and photochemical lifetimes of key molecules in extended atmospheres \citep{Tsai2021-ag, Hu2021-jj, Yu2021-qb, Wogan2024-af, Huang2024-ni}. The atmospheric presence of highly water soluble molecules such as \ce{NH3} or HCN could point to a dry or non existent surface, while the absence of these molecules, and relative abundances of \ce{CH4}, \ce{CO2}, \ce{H2O}, and \ce{C2H6} could suggest the planet is Hycean in nature. Other techniques such as comparing the \ce{CO2} to \ce{CH4} ratio have also been proposed to probe the interior of planets in this size regime \citep{Yang2024-pv}. 

Examinations of temperate sub-Neptune planets have been among the first rounds of observations made with JWST, including: TOI-270d \citep{Holmberg2024-fh,Benneke2024}, LHS 1140b \citep{Biagini2024-ru, Damiano2024-wo}, TOI-732b \citep{Cabot2024-vf}. One well-known example is the planet K2-18b, a sub-Neptune target which was previously observed with the Hubble Space Telescope (HST), producing the first claimed water detection on a sub-Neptune size planet \citep{Benneke2019-qp}. However, this claim has been disputed and the \ce{H2O} features have been reinterpreted as \ce{CH4} \citep{Madhusudhan2023-yh}. JWST observations of K2-18b \citep{Madhusudhan2023-ib} reported the presence of both \ce{CO2} and \ce{CH4}. Coupled with the lack of CO and \ce{NH3}, the authors argue that this atmospheric configuration is indicative of a Hycean planet. Additionally, \citep{Madhusudhan2023-ib} reported a tentative detection of dimethylsulfide (DMS, (CH$_{3}$)$_{2}$S)), a candidate atmospheric biosignature, originally proposed by \citet{pilcher2003} and quantitatively studied in anoxic terrestrial atmospheres by \citet{Domagal-Goldman2011-ko}. However, this interpretation is far from unanimous, with challenges and alternative interpretations of the potential biosignature presented in the literature. These include the possibility that thick atmosphere scenarios may better fit the data and that \ce{CH4} absorption at the wavelength reported may overprint any potential DMS signal \citep{Wogan2024-af, Tsai2024-gw}.  Some of these alternative models would indicate that K2-18b may be too hot to support a Hycean-type atmosphere with proposed compositions ranging from magma ocean to thick greenhouse atmospheres \citep{Leconte2024-tj, Shorttle2024-vf}. These substitute theories have been challenged in turn, showing that there is yet little consensus on the nature of this planet \citep{Rigby2024-tq, Cooke2024-le} . Regardless of how further observations confirm the Hycean nature of K2-18b, the general question of biosignatures in Hycean worlds has been brought to the forefront of the scientific community and remains an interesting opportunity given the accessibility of these targets to JWST. For example, compositionally similar planets receiving lower incident stellar flux may be amenable to Hycean conditions even if K2-18b in particular does not host a temperate liquid water ocean.

\cite{Tsai2024-gw} used photochemical and spectral simulations to perform vertically integrated simulations of the survivability and detectability of biogenic sulfur gases (including candidate methylated biosignature DMS) in the atmosphere of K2-18b, as a stand-in for Hycean worlds in general. This study found that for DMS to reach detectable levels a biological production flux of $\sim$20 times the globally averaged modern Earth flux is necessary. The authors applied both a 1D and 2D photochemical model and found that there is little difference between the simulated DMS outcomes, indicating that there is sufficient horizontal mixing in this case to oppose accumulation on the tidally locked nightside. Simulated spectra suggest that for NIRSpec wavelengths, DMS features are difficult to disentangle from \ce{CH4} with a cleaner feature accessible at mid-infrared wavelengths centered near $\sim$10 $\mu$m. 

DMS is a possible biosignature on Hycean worlds \citep{Madhusudhan2023-ib, Tsai2024-gw} because the gas is formed via biological methylation of environmental substrates, a process which also generates other biogenic gases on Earth. Its origin is overwhelming biological on Earth. In addition to DMS, other methylated chalcogens (S, Se, Te), halogens (Cl, Br, I), and metal(loid) compounds have been examined or proposed as potential astronomical biosignatures \citep{Segura2005-jv, Leung2022-us, meadows2023feasibility, SchwietermanLeung2024}. Numerous clades of bacteria and algae are known produce these methylated gases (i.e. \cite{Dimmer2001-kf,Shibazaki2016-le, Redeker2000-kj}). Depending on host star and production rate (surface flux), some of these gases can accumulate to potentially detectable atmospheric levels, shown through previous photochemical and spectral simulations that primarily assume terrestrial compositions \citep{Segura2005-jv, Leung2022-us, meadows2023feasibility, Angerhausen2024-qq}.

 Detection of methylated gases on Earth-like targets may require a larger investment in telescope time, but the low false positive potential of these gases provides significant value from a potential observation, motivating them as spectral targets for follow up observations. While proposed as potential biosignatures on Hycean worlds \citep{Leung2022-us,Madhusudhan2023-yh}, these biogenic gases have not yet been evaluated using coupled photochemical and spectral simulations to quantify the detectability of biologically plausible production rates of methyl halides for Hycean planets. The extended hydrogen envelope of these planets will enhance feature size and potential detectability, increasing the biosignature detection potential on Hycean worlds. Here we use the methods developed in \cite{Tsai2024-gw} and applied to DMS in that work to explore methyl halides as biosignatures on Hycean exoplanets.

\section{Methods} \label{sec:methods} 
\begin{table}[]
\centering
\begin{tabular}{@{}lclc@{}}
\toprule
Gas & Flux & Source \\ 
 & (molec/cm$^2$/s) &  \\ \midrule

\ce{CH3Br} & 5.17 $\times$ $10^6$ & \cite{Yang2005-pt} \\ 
\ce{CH3Cl} & 3.04 $\times$ $10^8$ & \cite{Xiao2010-kd} \\
\ce{CH3I} & 5.51 $\times$ $10^6$ & \cite{Ziska2013-tz} \\
\bottomrule
\end{tabular}
\caption{ Globally averaged biological surface production fluxes of methyl halides included in this study. Values reported are from Earth science models for \ce{CH3Cl} and \ce{CH3Br} used to analyse the cycling of halogens, with high \ce{CH3Cl} values representing the elevated abundances of Cl compared to other halogens. The \ce{CH3I} data reflect oceanic measurements on Earth, and are extrapolated to a globally averaged local surface flux.
 }\label{table:fluxes}
\end{table}

\subsection{VULCAN photochemical model} 
\begin{figure*}[tbh]

\subfigure[]{
\includegraphics[width=0.45\textwidth]{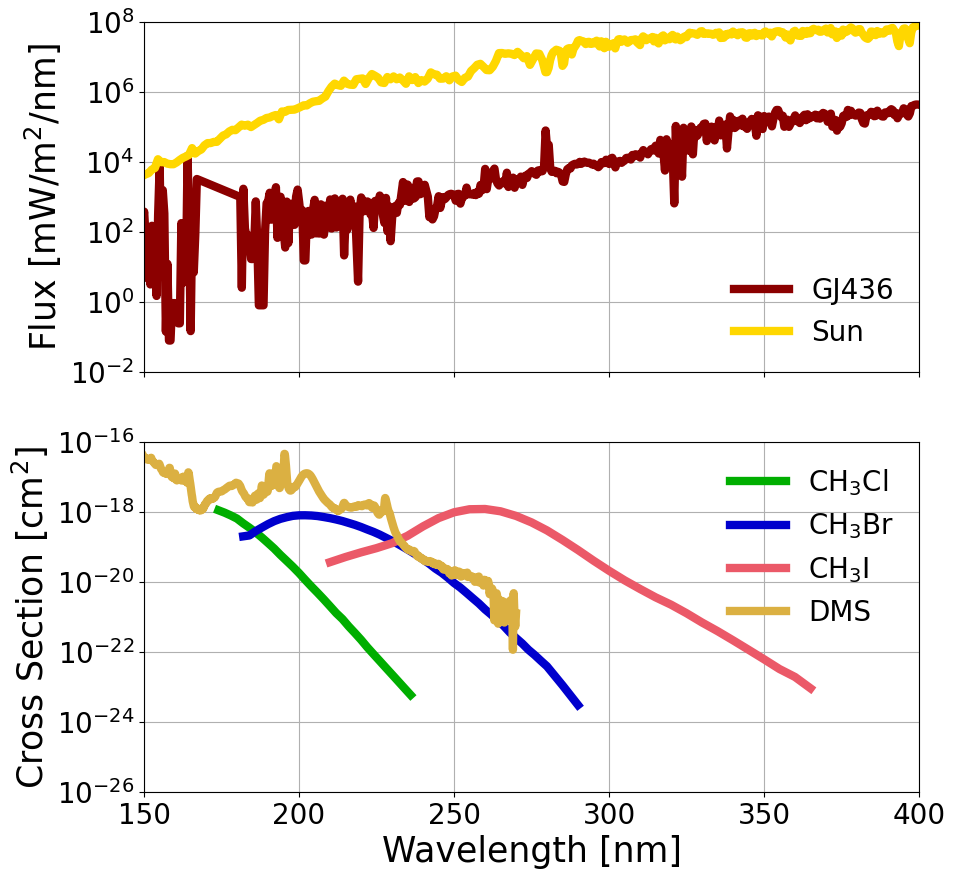}}
\centering
\subfigure[]{
\includegraphics[width=0.45\textwidth]{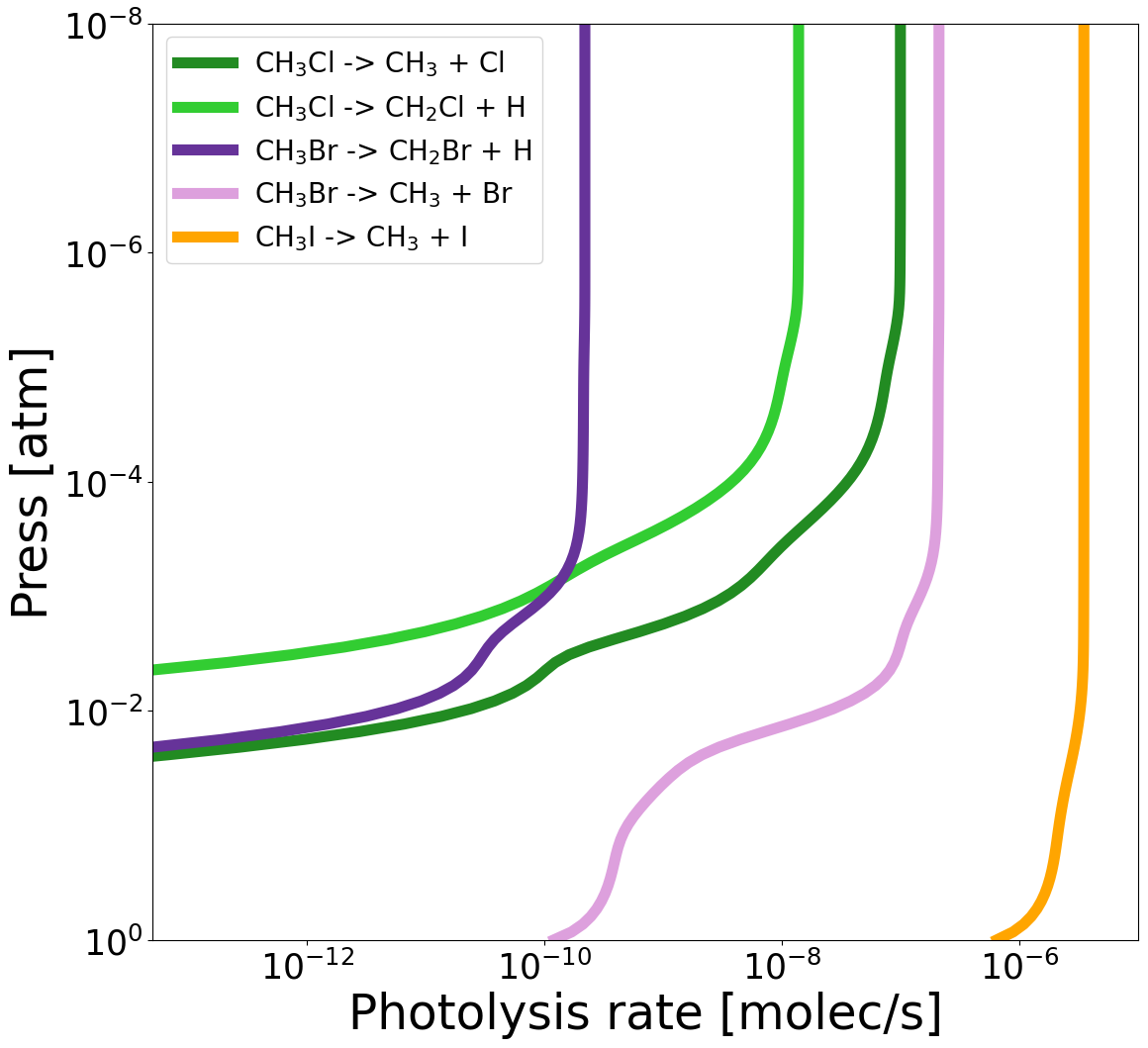}}
\centering

\subfigure[]{
  \includegraphics[width=0.45\textwidth]{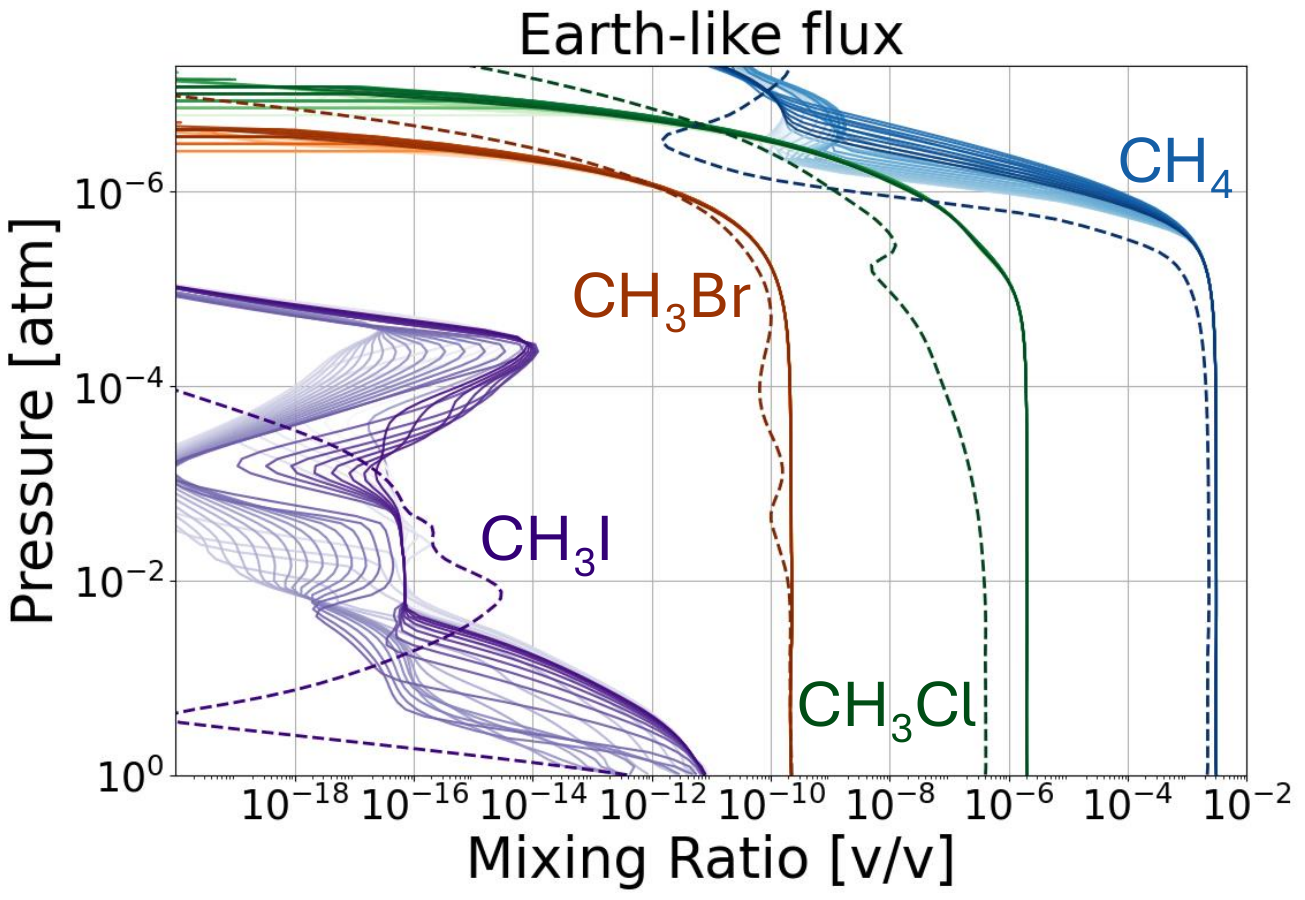}}
\subfigure[]{
   \includegraphics[width=0.45\textwidth]{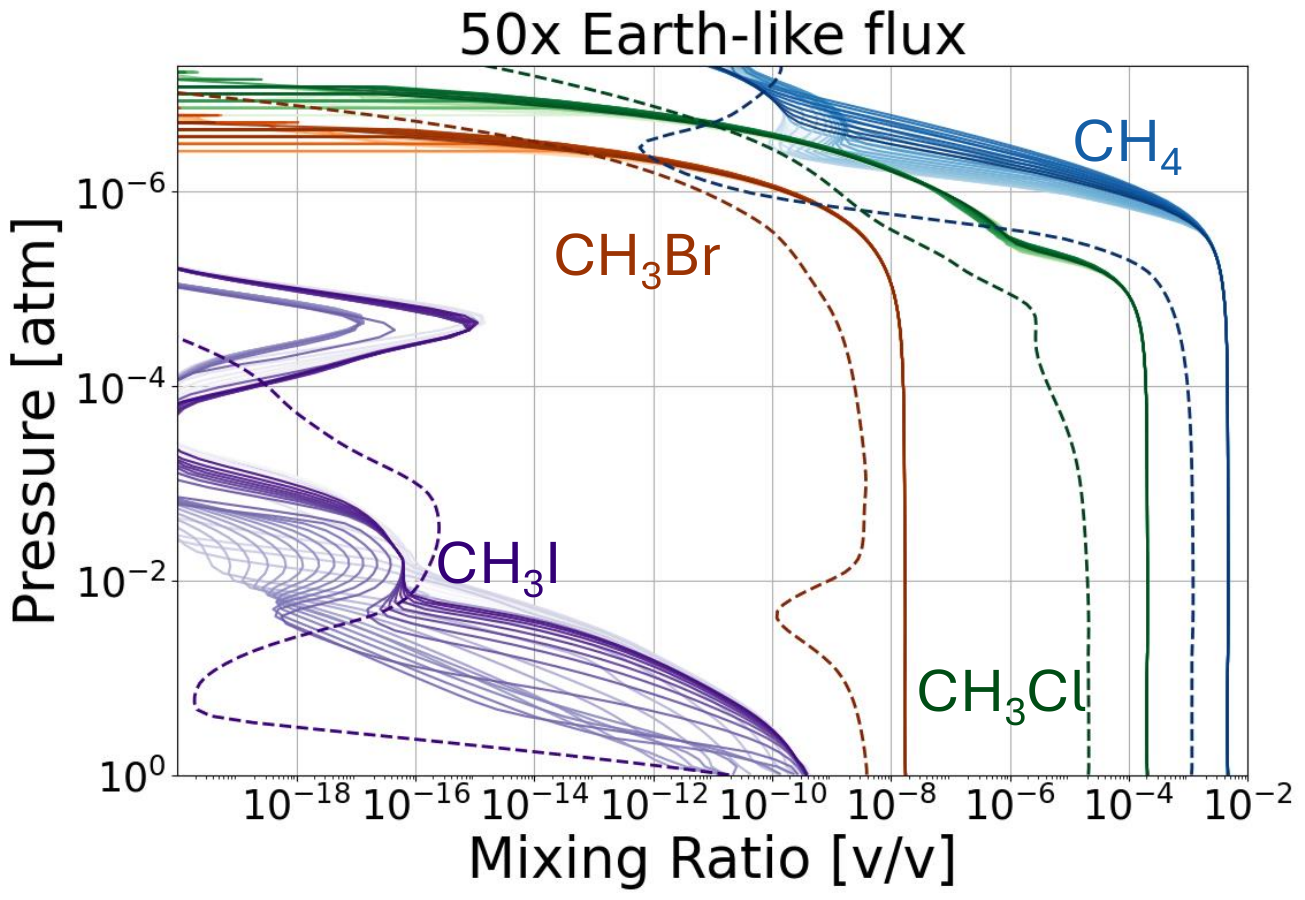}}

\caption{(a) Comparison of stellar surface spectra for both the Sun and GJ 436, used here as a stand in for K2-18, an M2.8V star (top) and photochemical cross sections for \ce{CH3X} gases and DMS (bottom). (b) Comparison of photolysis for \ce{CH3X} gases for Earth-like biological production flux levels simulated using K2-18b parameters. (c) Comparison of mixing ratio profiles for 1D (dashed) and 2D (solid, shaded by longitude) simulations using Earth-like flux levels. Modest enhancements of \ce{CH3Cl} are shown, with similar ratios of the other gases. Largest mixing ratio range is seen for \ce{CH3I} which shows some longitudinal dependency. (d) Mixing ratios similar to (c) for 50x Earth-like flux, around where fluxes becomes potentially spectrally detectable. Enhancements of 5-10x are seen for all gases compared to 1D results.}  
\label{fig:xsec}
\end{figure*}

We use the \texttt{VULCAN} photochemical code to model the potential accumulation of methyl halide biosignatures in Hycean atmospheres for a range of flux (surface production rate) conditions. \texttt{VULCAN} has been validated against a variety of planetary types including Earth, Hot Jupiters, and temperate sub-Neptune planets \citep{Tsai2021-ag, Tsai2021-sk, Huang2024-ni}. We adopt the Hycean boundary conditions used in \citep{Tsai2024-gw}, including the pressure-temperature profile generated for K2-18b. We utilize the same spectrum of GJ 436 and scaled solar spectra. For this study, we incorporate closed loop reaction networks for methyl chloride (\ce{CH3Cl}), methyl bromide (\ce{CH3Br}), and methyl iodide (\ce{CH3I}). This expansion totals 444 reactions including 38 photodissociation reactions as well as the necessary thermodynamical data and photochemical cross sections. See Figure \ref{fig:xsec} for a comparison of the photochemical cross sections used here, alongside the stellar spectra. Cross sectional data for \ce{(CH3)2S} are also included for comparison to previous Hycean studies (i.e., \cite{Tsai2024-gw}). We adopt the pressure-temperature profile from \cite{Tsai2024-gw} for a K2-18b-like Hycean planet without biological sources.

Building on the S-N-C-H-O photochemical network with DMS and DMDS in \cite{Tsai2024-gw}, we extend the reaction network to incorporate closed loop methyl halide reactions. The version of this code incorporating the halogen chemistry is available on Github\footnote{\url{https://github.com/MichaelaLeung/VULCAN_CH3X}} and our full boundary conditions are reported in Appendix \ref{appendix:bound}. We adopt the methyl halide (\ce{CH3X}, X = Cl, Br, I) biogenic surface fluxes from \cite{Leung2022-us}, with the addition of a \ce{CH3I} flux. We consider a range of gas fluxes up to 1000x the globally averaged flux on the Earth, a reasonable assumption given the known high spatial and temporal variability of the gas production (see \cite{Leung2022-us} for further discussion of highly productive organisms and environments). Marine ecosystems on Earth contribute to the global methyl halide flux \citep[i.e.][]{Moore2003-rg, Xiao2010-kd} so global ocean environments such as Hycean worlds could plausibly yield  biological production fluxes sufficient to generate atmospheric signals . Alternative evolutionary pathways could also lead to enhanced biosignature production on warm Hycean planets \citep{Mitchell2025-gr}.

Figure \ref{fig:xsec}cd show the mixing ratio profiles, simulated using both 2D \citep{Tsai2024-2D} and 1D photochemical models. The 2D results, shown in solid lines, shaded based on longitude, show moderate enhancements versus those modeled in 1D, especially for \ce{CH3Cl} and for the 50x biological production flux case. These results are consistent with \cite{Tsai2024-gw}, finding that horizontal transport can oppose extreme nightside accumulation, though an enhancement of a factor of several is possible. This effect may mean that the 1D simulations explored in depth here underestimate the potential accumulation of these gases for tidally locked Hycean planets and that detection of methyl halides may be easier than our results suggest. To understand the difference in mixing ratios of \ce{CH3Cl}, \ce{CH3Br}, and \ce{CH3I}, we compared the photolysis profiles generated from \texttt{VULCAN}. Figure \ref{fig:xsec}b, shows that, in addition to assumed lower surface biological fluxes as shown in Table \ref{table:fluxes}, \ce{CH3Br} and \ce{CH3I} are photolyzed much more rapidly than \ce{CH3Cl}, accounting for their lower relative enhancement in the simulated K2-18b atmosphere.

\subsection{Planetary Spectrum Generator}
To simulate the detection of methyl halide biosignatures, we use the Planetary Spectrum Generator (PSG; \cite{Villanueva2018-jt, Villanueva2022-lo}) to model transmission and emission spectra based on the atmospheric composition, planetary, stellar, and observational parameters. PSG was originally developed by \cite{Villanueva2018-jt} and has been has been used for a variety of solar system and exoplanet applications \citep[i.e.,][]{Pidhorodetska2020-ip, Suissa2020-sc, Liuzzi2021-hj, Villanueva2023-hb, Ranjan2023-hy, Eager-Nash2024-kr}. PSG uses correlated k-tables, and when necessary, line-by-line calculations to construct the atmospheric opacities. The input line lists for line-by-line calculations are from HITRAN and include measurements from 3 $\mu$m and longer wavelengths for methyl halides \citep{gordon2022hitran2020}. We simulate the noise reduction from multiple transits by dividing the noise by the square root of the number of transits being combined. 

Here we use the NIRSpec-PRISM and MIRI-LRS instrumental templates to simulate observations with JWST, including simulated multi source noise to determine the number of transits necessary to detect simulated features at 3 and 5$\sigma$ confidence. To calculate the signal-to-noise ratio (S/N), we determine the size of the feature by subtracting off the atmospheric continuum without the gas present. Then we use the simulated noise to find the S/N ratio. We determine the number of transits necessary to detect the feature by dividing the desired confidence by the square root of the binned signal to noise ratio as in \cite{Pidhorodetska2020-ip}. This method is common for first-pass observational analyses \citep[i.e.][]{Lustig-Yaeger2019-ke,Bixel2021-cx, Tsai2024-gw} and has been shown to strongly correlate with retrieval-based calculations \citep{Angerhausen2024-qq} . We bin the spectrum to R=30.

\section{Photochemical Results} \label{sec:pc_results}

\begin{figure*}[bth]
  \centering
  \includegraphics[width=0.93\textwidth]{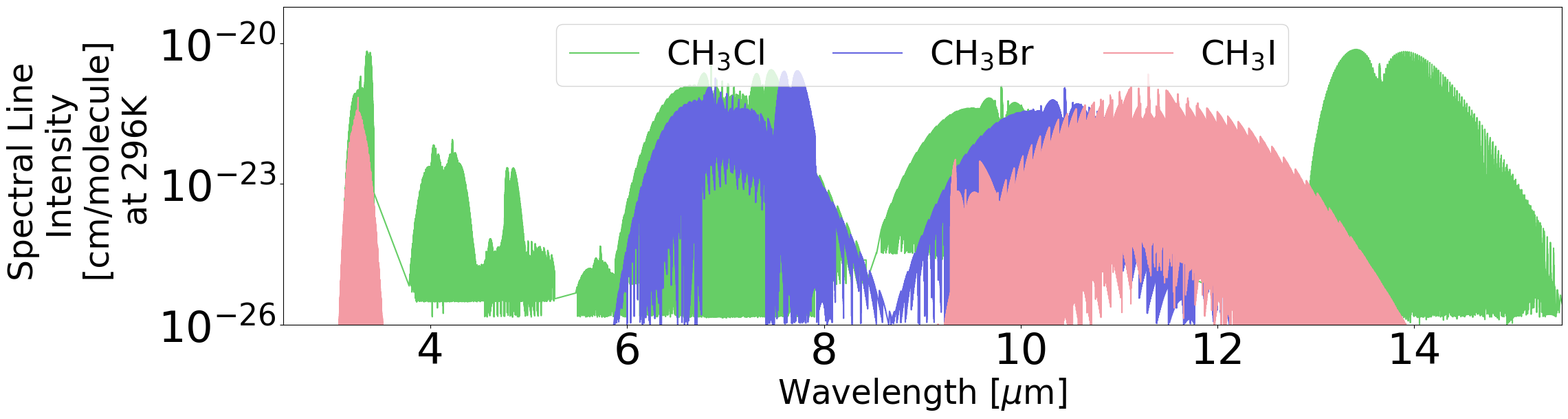}

  \includegraphics[width=0.95\textwidth]{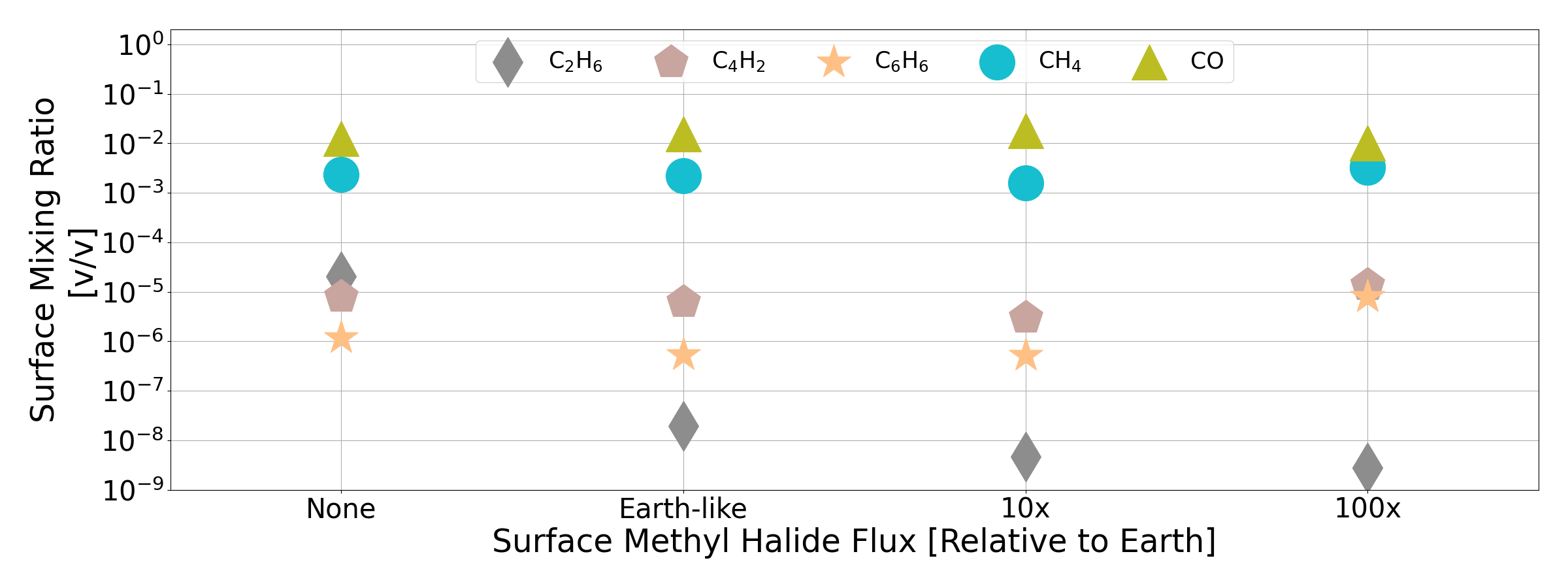}

  \includegraphics[width=0.95\textwidth]{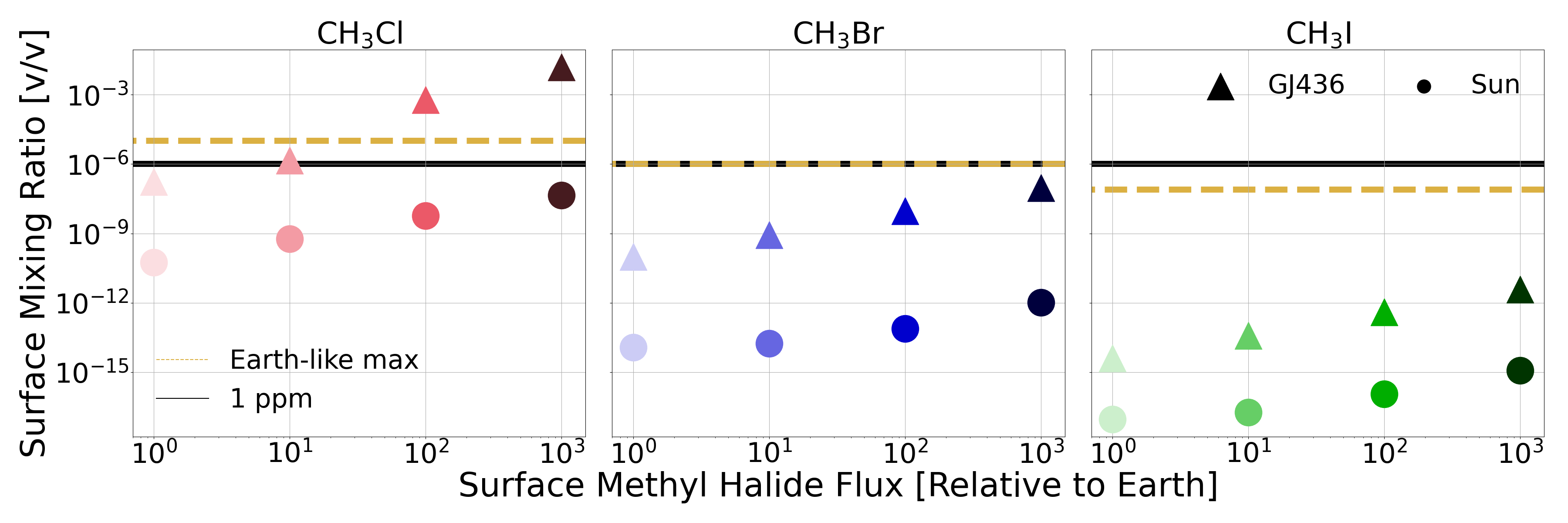}
  \caption{
  Top: Methyl halide opacities used in spectral simulations. Data sourced from the HITRAN database \cite{gordon2022hitran2020}. 
  Middle: Comparison of haze precursor molecules in Hycean atmospheres for a variety of methyl halide flux levels. These gases are investigated here because \cite{Tsai2024-gw} showed that that CO and \ce{C2H6} are highly responsive to DMS levels, with CO showing a large drop off and \ce{C2H6} a large increase as the organosulfur gas level increases. However, this trend does not hold here due to the rates of methyl halide reformation consuming the methyl radical and preventing further downstream chemical impacts. This is especially apparent in the large drop in \ce{C2H6} levels between the ``no Methyl Halide flux" case and the ``Earth-like Flux" .  
  Bottom: Comparison of average volume mixing ratios of methyl halide gases in Hycean-type atmospheres for the Sun and GJ 436, used here as a well characterized test case for a relatively inactive M2 star like K2-18. The triangles show the GJ 436 stellar environment and circle markers the solar environment. The difference in accumulation level for \ce{CH3Cl} is smaller in comparison to the other methyl halides, because \ce{CH3Cl} has less opacity in the NUV and is less affected by the increased brightness of the Sun at these shorter wavelengths (see Figure \ref{fig:xsec}). One part-per-million is shown with the black horizontal line as an approximate threshold for spectral relevance. Dashed yellow horizontal lines indicate mixing ratios reported for the most productive biological production flux scenario (1000x globally averaged) under modern Earth-like (\ce{O2}-rich) bulk conditions in \cite{Leung2022-us}.}
  \label{fig:phot_halo_comp}
\end{figure*}

The atmospheric accumulation of \ce{CH3Cl} in Hycean planets is a strong function of surface biological production flux, reaching ppm levels for the M dwarf host case at modest fluxes of 1-20 times Earth's global average, and percent levels at the highest productivity scenarios (1000x Earth flux; comparable to those found in highly productive environments like salt marshes). The production is above linear with the atmospheric concentration increasing 100x when the biological production flux is increased from 10x Earth-like to 100x Earth-like. As expected, the accumulation potential for Sun-like hosts is smaller, by about 2-3 orders of magnitude for each production rate scenario. This significant contrast is due to the reduced photolysis of methyl halides for M dwarf hosts. The reduction in photolysis rate is 4 orders of magnitude or greater for each methyl halide pathway. The effect of the stellar-driven photochemistry can be seen in Figure \ref{fig:phot_halo_comp}. Unlike \cite{Tsai2024-gw}, we do not find a significant impact from methyl halide fluxes on CO or \ce{C2H6} production. Methyl halides trend towards reformation (i.e. \ce{CH3} + Cl $\longrightarrow$ \ce{CH3Cl}) after photolysis, whereas there are fewer known pathways for DMS to do the same, resulting in fewer downstream products from \ce{CH3X} destruction. 

We also explore the impact of this planetary and stellar environment on other methyl halides. CH$_{3}$Br [CH$_{3}$I] present similar mixing ratios as recorded for modern Earth-like bulk atmospheres in \cite{Leung2022-us}, reaching close to ppm [10s of ppt] levels for maximum productivity cases. The lower build up of these gases, in comparison to \ce{CH3Cl}, can be attributed to higher photolysis rates of these molecules in the absence of \ce{O2}/\ce{O3} shielding. Additionally, these gases have lower biological production fluxes on Earth, and their peak photodissociation wavelengths intersect with a higher flux part of an M dwarf spectrum (see Figure \ref{fig:xsec}). For the simulations using the Sun as the host star , there is lower atmospheric accumulation, due to the increased photolysis resulting from enhanced total UV flux. 
 
\begin{figure*}[bth]
  \centering
  \includegraphics[width=0.95\textwidth]{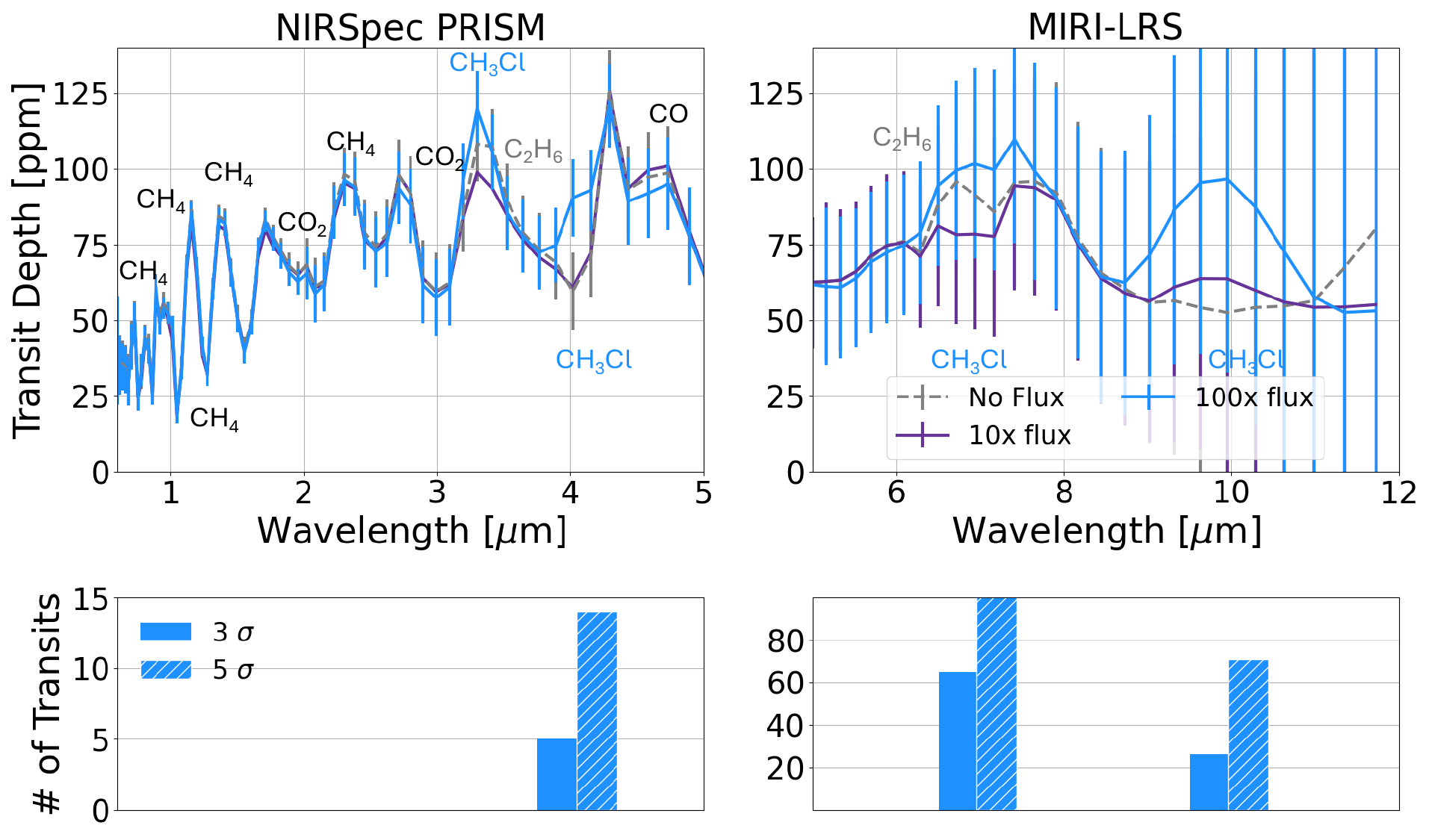}
  \caption{Top Left: Simulated observations using JWST NIRSpec-PRISM instrument. Optimal detection for \ce{CH3X} gases in 5 transits (3$\sigma$, 14 for 5$\sigma$) at the 4.0 $\mu$m \ce{CH3Cl} feature which does not have strong \ce{CH4} interference, as seen for other \ce{CH3Cl} features in this spectral range (i.e., 3.3 $\mu$m). The "\ce{CH3X}" opacity originates from \ce{CH3Cl} with the other gases not contributing to the spectrum at an observable level. Error bars are shown for 5 transits. 
  Top Right: Simulated observations using JWST MIRI-LRS instrument. Both features are more difficult to detect due to instrumental constraints, but detection is possible for the 10 $\mu$m band in tens of transits. For the "No Flux" scenario, \ce{C2H6} features are present near 3.3 $\mu$m, between 6-8 $\mu$m and beyond 12 $\mu$m, confounding the detection of methyl halide gases in these ranges. The presence of methyl halide gases at biological production levels appears to suppress \ce{C2H6} abundances, see Section \ref{sec:dis} for further discussion. Concentrations of other gases such as CO and \ce{H2CO} are also reduced when high levels of methyl radicals are introduced. Error bars shown are for 5 transits. 
  Bottom: Bar chart comparing number of transits necessary to detect each feature based on simulated noise for two confidence levels. } 
  \label{fig:spec_halo_comp}
\end{figure*}

\section{Atmospheric Detection} \label{sec:spec_results} 
\subsection{Using JWST}

Simulated spectra including noise parameters for NIRSpec PRISM \& MIRI instruments suggest that detection of combined \ce{CH3X} features, for the range of surface biological production fluxes considered here, may be possible in 5-14 transits of K2-18b-like planets, for optimistic biological production  flux levels. We use the system and planetary parameters for K2-18b throughout the spectral simulations as a test case planet, including in transit. For the 1000x globally averaged Earth biological production flux cases, the high levels of accumulated \ce{CH3Cl} increase the mean molecular weight (MMW) of the atmosphere enough to change the overall continuum due to a reduction in scale height. Here, we focus on the flux cases (base, 10x, 100x) which maintain a MMW \textless 4, since the inflated light atmosphere lends an observational advantage to the planet. For the highest biological production flux case, the MMW increases to 4.5 from 3.9 with the base \ce{CH3X} flux.  While our photochemical experiments consider all methyl halide gases, using those photochemical profiles to generate synthetic planetary spectra reveals that only \ce{CH3Cl} is sufficiently abundant to contribute atmospheric features for the flux cases considered here. \ce{CH3Br} and \ce{CH3I} opacities are considered in the radiative transfer model, but in practice have no impact on the spectrum due to their low predicted abundance. Hereafter, we refer to the \ce{CH3Cl} spectral features only. 

Our simulated transmission spectroscopy observations, generated using PSG, for the NIRSpec PRISM instrument suggest that for the 100x [50x] globally averaged Earth biological production flux , it is possible to detect \ce{CH3Cl} at 4.0 um with 3$\sigma$ confidence in 5 [12] transits of K2-18b. The other main "\ce{CH3X}" feature in this area is also \ce{CH3Cl} at 3.3 $\mu$m, however this feature overlaps strongly with methane absorption features so a diagnostic detection would be difficult, especially given the simultaneous elevated production of \ce{CH4} from the photochemical processing of \ce{CH3X} molecules. Other potential \ce{CH3X} features in this wavelength range are currently not quantitatively measured for \ce{CH3Br} and \ce{CH3I}. 

In the mid-infrared, the \ce{CH3X} feature at 10 $\mu$m, also dominated by \ce{CH3Cl}, is the best candidate, requiring 26 transits for 100x biological production flux at a 3$\sigma$ detection level, calculated based on the MIRI-LRS instrument. There are \ce{CH3Br} and \ce{CH3I} features within this region, but sensitivity tests confirm that they do not generate detectable features for the conditions and parameters simulated in this work. \ce{CH3Cl} dominates the absorption features due to higher concentrations in the atmosphere. These detections of \ce{CH3Cl} would represent a considerable investment of telescope time, however if other potential biosignatures had been detected on an exoplanet, searching for a capstone \ce{CH3Cl} signature in the mid-infrared may be justifiable and necessary to aid in interpretation. 

\subsection{Observability with the LIFE telescope}\label{LIFE_obs}
\begin{figure*}[bth]
\includegraphics[width=0.95\textwidth]{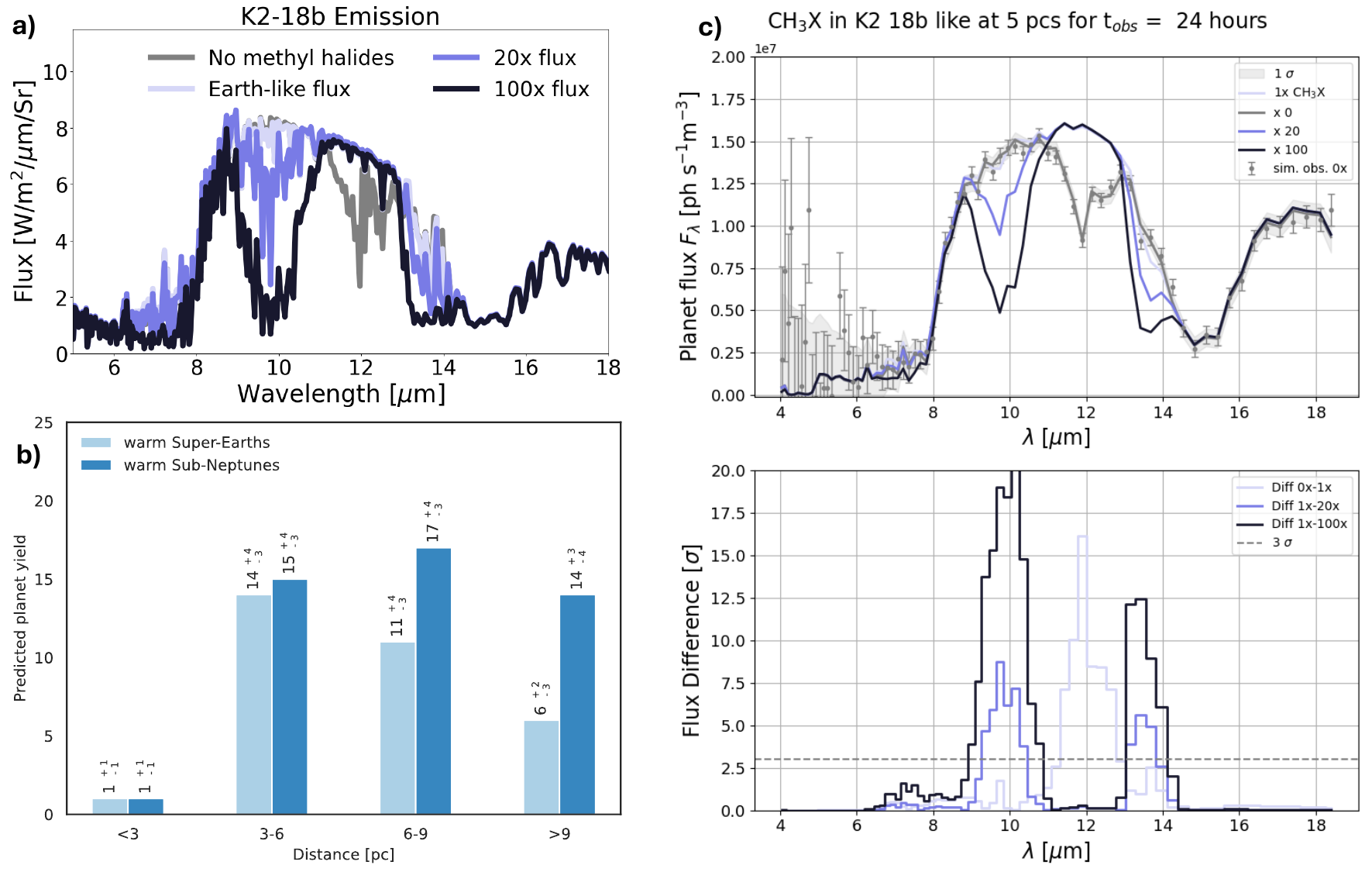}

  \caption{(a) Emission spectra of K2-18 b for various \ce{CH3X} biological production flux levels using photochemical profiles as input. The largest feature can be seen at 10 $\mu$m, with smaller features appearing at 7 and 13.7 $\mu$m. (b) Distance distribution of warm sub-Neptunes and super-Earths around M and FGK-type stars detectable with \textit{LIFE} in the current baseline setup. For details on the exoplanet classification see \citet{Kopparapu2018}. The 5 parsec distance assumed here is the typical distance for more than a dozen expected targets. (c) Detectability of various levels of CH$_3$X (primarily \ce{CH3Cl}) biological production fluxes in the emission spectrum of a ``K2 18b-like'' planet at 5 parsec, after 24 hours of observation with \textit{LIFE}. Top: planetary emission for atmospheres with and without various levels of CH$_3$Cl. The grey area represents the 1-$\sigma$ sensitivity; the grey error bars show an individual simulated observation. Bottom: Statistical significance of the detected differences between atmospheric models with various levels of CH$_3$X (see legend). } 
  \label{fig:emiss}
\end{figure*}

These potential biosignature molecules are also accessible via emission spectroscopy. Our results for simulated Hycean worlds show that the absorption features of the methylated gases in emission are comparable to other features such as \ce{H2O} and \ce{CO2}. Figure \ref{fig:emiss} compares the impact of different \ce{CH3Cl} biological production flux levels on emission spectra for K2-18b. In addition to the major \ce{CH3Cl} feature near 10 $\mu$m, there is additional \ce{CH3Cl} absorption between 6-8 $\mu$m, and longward centered at 13.7 $\mu$m, also noted by \cite{Rugheimer2013-je,Leung2022-us}. Both features are dominated by \ce{CH3Cl}, with minimal contributions by the other methyl halides. 

Following the same approach as in \cite{2023AsBio..23..183A} and \cite{Angerhausen2024-qq} we used LIFE\textsc{sim} \citep{2022A&A...664A..22D} to calculate the expected yields of K2-18b like planets with LIFE and produce synthetic observations of the outlined exoplanet cases with different biological production flux levels of the discussed species and also without them being present in their atmospheres. For the presented output spectra, LIFE\textsc{sim} is configured with the current \textit{LIFE} ``baseline" setup (Quantum efficiency 0.7,
  Throughput 0.05,
  Wavelength 4-18.5 $\mu$m,
  Spectral Resolution 50,
  Interferometric Baseline 10-100 m,
  Apertures Diameter 2m,
  Exozodi 3x local zodi).

Our analysis (see Fig. \ref{fig:emiss}b) shows that LIFE will be able to detect more than 70 warm super-Earth and sub-Neptune planets within 10 pcs. Based on this analysis, we choose 5 pcs as a typical distance for our simulations presented here. LIFE\textsc{sim} simulations of the expected signal-to-noise ratios (see Fig. \ref{fig:emiss}c) show that the various levels of CH$_3$X biological production fluxes discussed here will be detectable within only 24 hours of observations. This relatively small time requirement means that Hycean signals could be detected in the initial LIFE survey phase \citep[i.e][]{Quanz2022-oo} and may not require dedicated characterization time to detect molecular features of methyl halide for these targets.

\section{Discussion} \label{sec:dis}
The most favorable detection prospect for methylated gases in Hycean atmospheres would be under biological production flux conditions greater than 10x the globally averaged flux on the Earth, possibly through a greater radiation of the methylation pathways or high productivity in the marine environment. The most favorable wavelength for detection in transit is $\sim$ 4.0 $\mu$m where the NIRSpec PRISM instrument enables observations with lower noise. Our results suggest that to detect methylated gases on Hycean worlds with JWST requires a minimum of 5 transits, depending on the desired confidence level and observational wavelengths. 

As of this writing, JWST is currently the most capable telescope for potential biosignature detections on Hycean planets. However, NASA's planned next flagship telescope, currently called the Habitable Worlds Observatory, will be optimized for searching for signs of life via reflected light spectroscopy on temperate planets orbiting nearby stars. This instrument is predicted to span UV, visible, and NIR wavelengths up to 2 $\mu$m only \citep{NAP26141}. This direction for biosignature science, in combination with the results presented here, motivates further exploration of methylated gas features at shorter near-infrared wavelengths ($\lambda$ $\leq$ 2 $\mu$). Shorter wavelength features for \ce{CH4} in the NIR and optical hint that additional methyl halide absorption features may also exist, but wavelength specific opacities have not yet been measured at high fidelity. Further understanding of short wavelength \ce{CH3X} features through laboratory measurements is critical to constrain the applicability of these potential biosignatures to HWO. Complementarily, the LIFEsim results are highly favorable, showing that the time to detect \ce{CH3Cl} on a Hycean planet with thermal-IR emission spectroscopy is substantially lower than the time necessary to observe biosignature candidates on a terrestrial world, further motivating this additional next generation instrument \citep{Angerhausen2024-qq}. 

\ce{CH3Cl} is the main contributor to \ce{CH3X} features seen in this work, whereas in Earth-like (\ce{O2}-rich) atmospheres, \ce{CH3Br} can meaningfully contribute to the overall \ce{CH3X} spectral features \citep{Leung2022-us}. Altitude-dependent photolysis of \ce{CH3X} gases leads to this outcome with photolysis of \ce{CH3Br} and \ce{CH3I} occurring much closer to the surface (i.e., Figure \ref{fig:xsec}), resulting in both lower overall column densities and concentration of gases closer to the surface. Another reason that these molecules build up to lower levels is due to the lack of photochemical shielding, provided by \ce{O2}/\ce{O3} in an oxic environment. \ce{CH3Br} and \ce{CH3I} photolyze at longer wavelengths where the lack of shielding is more significant. Methyl halide gases are produced at different abundances based on the production environment \citep{Van_Pee2003-zn}; here we have used only the globally averaged ratios. In a global marine environment, the actual ratios may be different and \ce{CH3Br} or \ce{CH3I} could emerge as the dominant methyl halide, or provide meaningful spectral contribution . We note that there are limited abiotic sources for \ce{CH3Cl} and other methylated gases \citep[i.e.][]{Fayolle2017-ws, Hanni2024-ir, Sanz-Novo2025-ah}, but that the high destruction rates require a substantial source to overcome to detectable levels, unlikely to occur from modest yielding abiotic sources \cite{Leung2022-us}. 

Previous results have considered the role of DMS, also a methylated gas biosignature candidate in Hycean atmospheres \citep[e.g.][]{Tsai2024-gw}. In comparison to DMS, both \ce{CH3Cl} and \ce{CH3Br} have more modest surface biological production fluxes (i.e. Table \ref{table:fluxes}), which results in lower atmospheric accumulation, particularly for \ce{CH3Br}. We predict atmospheric accumulation of \ce{CH3Cl} to similar levels as DMS which is reasonable given the lower flux of the M2 host star at the relevant wavelengths for the \ce{CH3Cl} cross section. Studies of DMS also demonstrated that high concentrations of \ce{CH4} may obscure additional biosignature signals due to the dense opacity, particularly near 3.3 $\mu$m and plausible high accumulation. This effect is reduced for this study due to the efficient reformation of methyl halides, allowing for optimal observations of \ce{CH3Cl} at 4.0 $\mu$m . 

Our results are based on the best available absorption cross section measurements and spectral line list data. However, the necessary input absorption data has not been measured at high resolution or at all wavelengths relevant to exoplanetary studies. 
There is currently limited wavelength coverage for \ce{CH3Cl} photodissociation data \citep{Burkholder2020-lg}. If the cross section is larger than currently reported, there would be an increase in photolysis and decrease in mixing ratios. Constraining this uncertainty would enhance our confidence regarding the expected gas flux-abundance relationship of \ce{CH3Cl} on Hycean planets. 

In this work, we assume zero deposition velocity for \ce{CH3X} gases. In this scenario, surface sinks of the gases (i.e., biological consumption) are assumed to be of insignificant levels and the gases would exist at saturation levels in the ocean. With higher deposition velocities, the atmospheric accumulation would decrease as surface sinks would take up the gas instead. This would rely on a biological process which can consume large amounts of methylated gases. Similarly to \ce{CH4}, while there is biological and surface uptake of \ce{CH3X} gases on Earth, the rates are vastly exceeded by the rate of production \citep{Rhew2007-wk}, in part due to modest water solubility for methyl halides. \cite{Tsai2024-gw} explore the impact of changing deposition velocity on atmospheric accumulation of DMS, finding that surface deposition is a limiting control on mixing ratio for this gas, a trend which would likely hold for the similarly behaved \ce{CH3X} gases. 

Another factor not robustly explored here is the impact of the K$_{zz}$ parameter. We assume a K$_{zz}$ profile based on the GCM simulations performed for \cite{Tsai2024-gw}. Sensitivity tests suggest that higher K$_{zz}$ values may contribute to more atmospheric accumulation of methylated gases in the upper atmosphere. These higher abundances would be easier to detect in spectral observations. Our K$_{zz}$ values represent a conservative control on atmospheric accumulation where actual ability to detect may be easier if eddy diffusion throughout the atmosphere is greater. 

\section{Conclusion} 
Building on previous work examining methylated gases as biosignatures for Earth-like planets, we consider their possible observability on potentially habitable sub-Neptune planets with a hydrogen envelope sitting over an ocean, creating conventionally habitable conditions. Atmospheric accumulation of \ce{CH3Cl} in this atmosphere type, simulated using K2-18b as a potential type case, shows that methylated gases (particularly \ce{CH3Cl}) are well suited to build up in these environments and can easily reach ppm levels and even \% levels for optimistic biological production flux assumptions projections. Spectral simulations reveal that these gas features are potentially detectable in as few as 14 transits using JWST instruments or 24 hours using next generation space-based emission spectroscopy. These results support the use of methylated gases as corroborative ``capstone" biosignatures in the near future, should any promising targets be revealed through preliminary characterization. 

\begin{acknowledgments}
This work was supported by the NASA Interdisciplinary Consortia for Astrobiology Research (ICAR) program via the CHAMPs (Consortium on Habitability and Atmospheres of M-dwarf Planets) team with funding issued under grant no. 80NSSC23K1399, the Alternative Earths team with funding issued under grant no. 80NSSC21K0594, and the Virtual Planetary Laboratory (VPL) team with funding issue under grant no. 80NSSC23K1398. Computations were performed using the computer clusters and data storage resources of the UCR HPCC, which were funded by grants from NSF (MRI-2215705, MRI-1429826) and NIH (1S10OD016290-01A1). ML thanks the UC President's Dissertation Year Fellowship for partial funding of this work. The authors would also like to thank Vincent Kofman for valuable comments concerning the noise modeling simulations. 
\end{acknowledgments}

\vspace{5mm}

\software{ \href{https://matplotlib.org/}{matplotlib \citep{Hunter:2007}}, \href{https://numpy.org/}{numpy \citep{numpy}}, \href{https://www.python.org}{python \citep{python}},\href{https://psg.gsfc.nasa.gov/}{Planetary Spectrum Generator \citep{Villanueva2018-jt}}, \href{https://github.com/exoclime/VULCAN}{VULCAN \citep{Tsai2017-nw,Tsai2021-sk}}, \href{https://github.com/ACCarnall/spectres}{spectres \citep{carnall2017spectres}}}
Software citation information aggregated using \texttt{\href{https://www.tomwagg.com/software-citation-station/}{The Software Citation Station}} \citep{software-citation-station-paper, software-citation-station-zenodo}.
\clearpage

\appendix
\section{Photochemical Model Boundary Conditions} \label{appendix:bound}
Our gas flux boundary conditions, including deposition velocities, are supplied for ease of replication in Table \ref{tab:boundary}. 

\setcounter{table}{0}
\renewcommand{\thetable}{A}
\begin{table}[hbt]
\begin{tabular}{@{}lll@{}}
\toprule
Gas   & Flux (molec/cm$^2$/s) & Deposition Velocity (cm/s) \\ \midrule
\ce{SO2} & 9 $\times$ 10$^9$  & 1.0            \\
\ce{H2S} & 2 $\times$ 10$^8$  & 0.015           \\
\ce{H2O2} & 0         & 1.0            \\
\ce{S}  & 0         & 1.0            \\
\ce{SO}  & 0         & 3.0 $\times$ 10$^{-4}$   \\
\ce{HSO} & 0         & 1.0            \\
\ce{CH3S} & 0         & 0.01            \\
\ce{COS} & 5.4 $\times$ 10$^7$ & 0.003           \\
\ce{CH3CH3} & 4.2 $\times$ 10$^9$ & 0.0            \\
\ce{CH3SH} & 8.3 $\times$ 10$^8$ & 0.0            \\
\ce{CS2} & 1.4 $\times$ 10$^7$ & 0.0            \\
\ce{CH4} & 7.0 $\times$ 10$^{10}$ & 0.0            \\
\ce{CH3Br} & 5.17 $\times$ 10$^6$ & 0.0            \\
\ce{CH3Cl} & 3.04 $\times$ 10$^8$ & 0.0            \\
\ce{CH3I} & 5.51 $\times$ 10$^6$ & 0.0            \\
\ce{ClO } & 0         & 0.5            \\
\ce{HOCl} & 0         & 0.5            \\
\ce{Cl2} & 0         & 1.0            \\
\ce{ClONO2} & 0         & 0.5            \\
\ce{CH2ClO2} & 0         & 1.0            \\
\ce{HCl}  & 1.0$\times$ 10$^8$ & 0.2            \\
\ce{Cl}  & 0         & 1.0            \\
\ce{HClO4}& 0         & 0.2            \\
\ce{Br}  & 1.51$\times$ 10$^8$ & 1.0            \\
\ce{BrO}  & 0         & 0.5            \\
\ce{HBr}  & 1.0$\times$ 10$^6$ & 0.75            \\
\ce{Br2}  & 7.59$\times$ 10$^6$ & 0.01            \\
\ce{HOBr}  & 0        & 0.35            \\
\ce{CH2Br}  & 0        & 1.0            \\
\ce{IO}  & 1.0$\times$ 10$^{7}$  & 0.0            \\
\ce{I2}  & 0        & 1.0            \\
\ce{HI}  & 3.2$\times$ 10$^3$  & 0.0            \\
\ce{CH3Cl}  & various  & 0.0            \\
\ce{CH3Br}  & various  & 0.0            \\
\ce{CH3I}  & various  & 0.0            \\

 \bottomrule
\end{tabular}
\label{tab:boundary}
\caption{Boundary conditions for input in VULCAN photochemical model.}

\end{table}
\clearpage



\bibliography{sample631}{}

\begin{thebibliography}{}
\expandafter\ifx\csname natexlab\endcsname\relax\def\natexlab#1{#1}\fi
\providecommand{\url}[1]{\href{#1}{#1}}
\providecommand{\dodoi}[1]{doi:~\href{http://doi.org/#1}{\nolinkurl{#1}}}
\providecommand{\doeprint}[1]{\href{http://ascl.net/#1}{\nolinkurl{http://ascl.net/#1}}}
\providecommand{\doarXiv}[1]{\href{https://arxiv.org/abs/#1}{\nolinkurl{https://arxiv.org/abs/#1}}}

\bibitem[{Angerhausen {et~al.}(2024)Angerhausen, Pidhorodetska, \& {others}}]{Angerhausen2024-qq}
Angerhausen, D., Pidhorodetska, D., \& {others}. 2024, Astron. J.

\bibitem[{{Angerhausen} {et~al.}(2023){Angerhausen}, {Ottiger}, {Dannert}, {Miguel}, {Sousa-Silva}, {Kammerer}, {Menti}, {Alei}, {Konrad}, {Wang}, {Quanz}, \& {LIFE Collaboration}}]{2023AsBio..23..183A}
{Angerhausen}, D., {Ottiger}, M., {Dannert}, F., {et~al.} 2023, Astrobiology, 23, 183, \dodoi{10.1089/ast.2022.0010}

\bibitem[{Benneke {et~al.}(2019)Benneke, Wong, Piaulet, Knutson, Lothringer, Morley, Crossfield, Gao, Greene, Dressing, Dragomir, Howard, McCullough, Kempton, Fortney, \& Fraine}]{Benneke2019-qp}
Benneke, B., Wong, I., Piaulet, C., {et~al.} 2019, Astrophys. J. Lett., 887, L14, \dodoi{10.3847/2041-8213/ab59dc}

\bibitem[{{Benneke} {et~al.}(2024){Benneke}, {Roy}, {Coulombe}, {Radica}, {Piaulet}, {Ahrer}, {Pierrehumbert}, {Krissansen-Totton}, {Schlichting}, {Hu}, {Yang}, {Christie}, {Thorngren}, {Young}, {Pelletier}, {Knutson}, {Miguel}, {Evans-Soma}, {Dorn}, {Gagnebin}, {Fortney}, {Komacek}, {MacDonald}, {Raul}, {Cloutier}, {Acuna}, {Lafreni{\`e}re}, {Cadieux}, {Doyon}, {Welbanks}, \& {Allart}}]{Benneke2024}
{Benneke}, B., {Roy}, P.-A., {Coulombe}, L.-P., {et~al.} 2024, arXiv e-prints, arXiv:2403.03325, \dodoi{10.48550/arXiv.2403.03325}

\bibitem[{Bergsten {et~al.}(2022)Bergsten, Pascucci, Mulders, Fernandes, \& Koskinen}]{Bergsten2022-vw}
Bergsten, G.~J., Pascucci, I., Mulders, G.~D., Fernandes, R.~B., \& Koskinen, T.~T. 2022, AJS, 164, 190, \dodoi{10.3847/1538-3881/ac8fea}

\bibitem[{Biagini {et~al.}(2024)Biagini, Cracchiolo, Petralia, Maldonado, Di~Maio, \& Micela}]{Biagini2024-ru}
Biagini, A., Cracchiolo, G., Petralia, A., {et~al.} 2024.
\newblock \doarXiv{2403.20285}

\bibitem[{Bixel \& Apai(2021)}]{Bixel2021-cx}
Bixel, A., \& Apai, D. 2021, Astron. J., 161, 228, \dodoi{10.3847/1538-3881/abe042}

\bibitem[{Burkholder {et~al.}(2020)Burkholder, Sander, Abbatt, Barker, Cappa, Crounse, Dibble, Huie, Kolb, Kurylo, L, Percival, Wilmouth, Wine, \& Orkin}]{Burkholder2020-lg}
Burkholder, J.~B., Sander, S.~P., Abbatt, J. P.~D., {et~al.} 2020, Chemical Kinetics and Photochemical Data for Use in Atmospheric Studies Evaluation Number 19, Tech. rep., \dodoi{10.1002/kin.550171010}

\bibitem[{Cabot {et~al.}(2024)Cabot, Madhusudhan, Constantinou, Valencia, Vos, Masseron, \& Cheverall}]{Cabot2024-vf}
Cabot, S. H.~C., Madhusudhan, N., Constantinou, S., {et~al.} 2024, ApJL, 966, L10, \dodoi{10.3847/2041-8213/ad3828}

\bibitem[{Carnall(2017)}]{carnall2017spectres}
Carnall, A. 2017, arXiv preprint arXiv:1705.05165

\bibitem[{Claudi {et~al.}(2020)Claudi, Alei, Battistuzzi, Cocola, Erculiani, Pozzer, Salasnich, Simionato, Squicciarini, Poletto, \& La~Rocca}]{Claudi2020-xt}
Claudi, R., Alei, E., Battistuzzi, M., {et~al.} 2020, Life, 11, \dodoi{10.3390/life11010010}

\bibitem[{Cooke \& Madhusudhan(2024)}]{Cooke2024-le}
Cooke, G.~J., \& Madhusudhan, N. 2024, Astrophys. J., 977, 209, \dodoi{10.3847/1538-4357/ad8cda}

\bibitem[{Damiano {et~al.}(2024)Damiano, Bello-Arufe, Yang, \& Hu}]{Damiano2024-wo}
Damiano, M., Bello-Arufe, A., Yang, J., \& Hu, R. 2024, ApJL, 968, L22, \dodoi{10.3847/2041-8213/ad5204}

\bibitem[{{Dannert} {et~al.}(2022){Dannert}, {Ottiger}, {Quanz}, {Laugier}, {Fontanet}, {Gheorghe}, {Absil}, {Dandumont}, {Defr{\`e}re}, {Gasc{\'o}n}, {Glauser}, {Kammerer}, {Lichtenberg}, {Linz}, {Loicq}, \& {LIFE Collaboration}}]{2022A&A...664A..22D}
{Dannert}, F.~A., {Ottiger}, M., {Quanz}, S.~P., {et~al.} 2022, \aap, 664, A22, \dodoi{10.1051/0004-6361/202141958}

\bibitem[{Dimmer {et~al.}(2001)Dimmer, Simmonds, Nickless, \& Bassford}]{Dimmer2001-kf}
Dimmer, C.~H., Simmonds, P.~G., Nickless, G., \& Bassford, M.~R. 2001, Atmos. Environ., 35, 321, \dodoi{10.1016/S1352-2310(00)00151-5}

\bibitem[{Domagal-Goldman {et~al.}(2011)Domagal-Goldman, Meadows, Claire, \& Kasting}]{Domagal-Goldman2011-ko}
Domagal-Goldman, S.~D., Meadows, V.~S., Claire, M.~W., \& Kasting, J.~F. 2011, Astrobiology, 11, 419, \dodoi{10.1089/ast.2010.0509}

\bibitem[{Eager-Nash {et~al.}(2024)Eager-Nash, Daines, McDermott, Andrews, Grain, Bishop, Rogers, Smith, Khalek, Boxer, Mak, Ridgway, H{\'e}brard, Lambert, Lenton, \& Mayne}]{Eager-Nash2024-kr}
Eager-Nash, J.~K., Daines, S.~J., McDermott, J.~W., {et~al.} 2024, Mon. Not. R. Astron. Soc., 531, 468, \dodoi{10.1093/mnras/stae1142}

\bibitem[{Fayolle {et~al.}(2017)Fayolle, {\"O}berg, J{\o}rgensen, Altwegg, Calcutt, M{\"u}ller, Rubin, van~der Wiel, Bjerkeli, Bourke, Coutens, van Dishoeck, Drozdovskaya, Garrod, Ligterink, Persson, \& Wampfler}]{Fayolle2017-ws}
Fayolle, E.~C., {\"O}berg, K.~I., J{\o}rgensen, J.~K., {et~al.} 2017, Nature Astronomy, 1, 703, \dodoi{10.1038/s41550-017-0237-7}

\bibitem[{Gordon {et~al.}(2022)Gordon, Rothman, Hargreaves, Hashemi, Karlovets, Skinner, Conway, Hill, Kochanov, Tan, {et~al.}}]{gordon2022hitran2020}
Gordon, I.~E., Rothman, L.~S., Hargreaves, e.~R., {et~al.} 2022, Journal of quantitative spectroscopy and radiative transfer, 277, 107949, \dodoi{10.1016/j.jqsrt.2021.107949}

\bibitem[{Greiss {et~al.}(2012{\natexlab{a}})Greiss, Steeghs, G{\"a}nsicke, Mart{\'\i}n, Groot, Irwin, Gonz{\'a}lez-Solares, Greimel, Gentile~Fusillo, Still, \& {the KIS collaboration}}]{Greiss2012-bx}
Greiss, S., Steeghs, D. T.~H., G{\"a}nsicke, B.~T., {et~al.} 2012{\natexlab{a}}.
\newblock \doarXiv{1212.3613}

\bibitem[{Greiss {et~al.}(2012{\natexlab{b}})Greiss, Steeghs, G{\"a}nsicke, Mart{\'\i}n, Groot, Irwin, Gonz{\'a}lez-Solares, Greimel, Knigge, {\O}stensen, Verbeek, Drew, Drake, Jonker, Ripepi, Scaringi, Southworth, Still, Wright, Farnhill, van Haaften, \& Shah}]{Greiss2012-dj}
Greiss, S., Steeghs, D., G{\"a}nsicke, B.~T., {et~al.} 2012{\natexlab{b}}, AJS, 144, 24, \dodoi{10.1088/0004-6256/144/1/24}

\bibitem[{Guerrero {et~al.}(2021)Guerrero, Seager, Huang, Vanderburg, Soto, Mireles, Hesse, Fong, Glidden, Shporer, Latham, Collins, Quinn, Burt, Dragomir, Crossfield, Vanderspek, Fausnaugh, Burke, Ricker, Daylan, Essack, G{\"u}nther, Osborn, Pepper, Rowden, Sha, Villanueva, Yahalomi, Yu, Ballard, Batalha, Berardo, Chontos, Dittmann, Esquerdo, Mikal-Evans, Jayaraman, Krishnamurthy, Louie, Mehrle, Niraula, Rackham, Rodriguez, Rowden, Sousa-Silva, Watanabe, Wong, Zhan, Zivanovic, Christiansen, Ciardi, Swain, Lund, Mullally, Fleming, Rodriguez, Boyd, Quintana, Barclay, Col{\'o}n, Rinehart, Schlieder, Clampin, Jenkins, Twicken, Caldwell, Coughlin, Henze, Lissauer, Morris, Rose, Smith, Tenenbaum, Ting, Wohler, Bakos, Bean, Berta-Thompson, Bieryla, Bouma, Buchhave, Butler, Charbonneau, Doty, Ge, Holman, Howard, Kaltenegger, Kane, Kjeldsen, Kreidberg, Lin, Minsky, Narita, Paegert, P{\'a}l, Palle, Sasselov, Spencer, Sozzetti, Stassun, Torres, Udry, \& Winn}]{Guerrero2021-ox}
Guerrero, N.~M., Seager, S., Huang, C.~X., {et~al.} 2021, ApJS, 254, 39, \dodoi{10.3847/1538-4365/abefe1}

\bibitem[{Harris {et~al.}(2020)Harris, Millman, van~der Walt, Gommers, Virtanen, Cournapeau, Wieser, Taylor, Berg, Smith, Kern, Picus, Hoyer, van Kerkwijk, Brett, Haldane, del R{\'{i}}o, Wiebe, Peterson, G{\'{e}}rard-Marchant, Sheppard, Reddy, Weckesser, Abbasi, Gohlke, \& Oliphant}]{numpy}
Harris, C.~R., Millman, K.~J., van~der Walt, S.~J., {et~al.} 2020, Nature, 585, 357, \dodoi{10.1038/s41586-020-2649-2}

\bibitem[{Holmberg \& Madhusudhan(2024)}]{Holmberg2024-fh}
Holmberg, M., \& Madhusudhan, N. 2024.
\newblock \doarXiv{2403.03244}

\bibitem[{Hu {et~al.}(2021)Hu, Damiano, Scheucher, Kite, Seager, \& Rauer}]{Hu2021-jj}
Hu, R., Damiano, M., Scheucher, M., {et~al.} 2021, ApJL, 921, L8, \dodoi{10.3847/2041-8213/ac1f92}

\bibitem[{Hu {et~al.}(2019)Hu, Beichman, Brain, Chen, Damiano, Dawson, James~Friedson, Hasagawa, Howard, Johnson, Kataria, Kidd, Kite, Knutson, Lyra, Mischna, Planavsky, Reinhard, Schlichting, Seager, Sotin, Swain, Turner, West, Yung, \& Zellem}]{Hu2019-qu}
Hu, R., Beichman, C.~A., Brain, D., {et~al.} 2019.
\newblock \doarXiv{1903.05258}

\bibitem[{Huang {et~al.}(2024)Huang, Yu, Tsai, Moses, Ohno, Krissansen-Totton, Zhang, \& Fortney}]{Huang2024-ni}
Huang, Z., Yu, X., Tsai, S.-M., {et~al.} 2024, arXiv [astro-ph.EP], arXiv:2407.09009, \dodoi{10.48550/arXiv.2407.09009}

\bibitem[{Hunter(2007)}]{Hunter:2007}
Hunter, J.~D. 2007, Computing in Science \& Engineering, 9, 90, \dodoi{10.1109/MCSE.2007.55}

\bibitem[{Hänni {et~al.}(2024)Hänni, Altwegg, Combi, Fuselier, De~Keyser, Ligterink, Rubin, \& Wampfler}]{Hanni2024-ir}
Hänni, N., Altwegg, K., Combi, M., {et~al.} 2024, Astrophys. J., \dodoi{10.3847/1538-4357/ad8565}

\bibitem[{{Kopparapu} {et~al.}(2018){Kopparapu}, {H{\'e}brard}, {Belikov}, {Batalha}, {Mulders}, {Stark}, {Teal}, {Domagal-Goldman}, \& {Mandell}}]{Kopparapu2018}
{Kopparapu}, R.~K., {H{\'e}brard}, E., {Belikov}, R., {et~al.} 2018, \apj, 856, 122, \dodoi{10.3847/1538-4357/aab205}

\bibitem[{Leconte {et~al.}(2024)Leconte, Spiga, Cl{\'e}ment, Guerlet, Selsis, Milcareck, Cavali{\'e}, Moreno, Lellouch, Carri{\'o}n-Gonz{\'a}lez, Charnay, \& Lef{\`e}vre}]{Leconte2024-tj}
Leconte, J., Spiga, A., Cl{\'e}ment, N., {et~al.} 2024.
\newblock \doarXiv{2401.06608}

\bibitem[{Leung {et~al.}(2022)Leung, Schwieterman, Parenteau, \& Fauchez}]{Leung2022-us}
Leung, M., Schwieterman, E.~W., Parenteau, M.~N., \& Fauchez, T.~J. 2022, ApJ, 938, 6, \dodoi{10.3847/1538-4357/ac8799}

\bibitem[{Liuzzi {et~al.}(2021)Liuzzi, Villanueva, Viscardy, M{\`e}ge, Crismani, Aoki, Gurgurewicz, Tesson, Mumma, Smith, Faggi, Kofman, Knutsen, Daerden, Neary, Schmidt, Trompet, Erwin, Robert, Thomas, Ristic, Bellucci, Lopez-Moreno, Patel, \& Vandaele}]{Liuzzi2021-hj}
Liuzzi, G., Villanueva, G.~L., Viscardy, S., {et~al.} 2021, Geophys. Res. Lett., 48, \dodoi{10.1029/2021gl092650}

\bibitem[{Lustig-Yaeger {et~al.}(2019)Lustig-Yaeger, Meadows, \& Lincowski}]{Lustig-Yaeger2019-ke}
Lustig-Yaeger, J., Meadows, V.~S., \& Lincowski, A.~P. 2019, AJS, 158, 27, \dodoi{10.3847/1538-3881/ab21e0}

\bibitem[{Madhusudhan {et~al.}(2023{\natexlab{a}})Madhusudhan, Moses, Rigby, \& Barrier}]{Madhusudhan2023-yh}
Madhusudhan, N., Moses, J.~I., Rigby, F., \& Barrier, E. 2023{\natexlab{a}}, Faraday Discuss., \dodoi{10.1039/d3fd00075c}

\bibitem[{Madhusudhan {et~al.}(2020)Madhusudhan, Nixon, Welbanks, Piette, \& Booth}]{Madhusudhan2020-kn}
Madhusudhan, N., Nixon, M.~C., Welbanks, L., Piette, A. A.~A., \& Booth, R.~A. 2020, ApJL, 891, L7, \dodoi{10.3847/2041-8213/ab7229}

\bibitem[{Madhusudhan {et~al.}(2021)Madhusudhan, Piette, \& Constantinou}]{Madhusudhan2021-io}
Madhusudhan, N., Piette, A. A.~A., \& Constantinou, S. 2021, Astrophys. J., 918, 1, \dodoi{10.3847/1538-4357/abfd9c}

\bibitem[{Madhusudhan {et~al.}(2023{\natexlab{b}})Madhusudhan, Sarkar, Constantinou, Holmberg, Piette, \& Moses}]{Madhusudhan2023-ib}
Madhusudhan, N., Sarkar, S., Constantinou, S., {et~al.} 2023{\natexlab{b}}, ApJL, 956, L13, \dodoi{10.3847/2041-8213/acf577}

\bibitem[{Meadows {et~al.}(2023)Meadows, Lincowski, \& Lustig-Yaeger}]{meadows2023feasibility}
Meadows, V.~S., Lincowski, A.~P., \& Lustig-Yaeger, J. 2023, The Planetary Science Journal, 4, 192, \dodoi{10.3847/PSJ/acf488}

\bibitem[{Mitchell \& Madhusudhan(2025)}]{Mitchell2025-gr}
Mitchell, E.~G., \& Madhusudhan, N. 2025, Mon. Not. R. Astron. Soc., staf094, \dodoi{10.1093/mnras/staf094}

\bibitem[{Moore(2003)}]{Moore2003-rg}
Moore, R.~M. 2003, in The Handbook of Environmental Chemistry, The handbook of environmental chemistry (Berlin, Heidelberg: Springer Berlin Heidelberg), 85--101, \dodoi{10.1007/b10449}

\bibitem[{{National Academies of Sciences Engineering and Medicine}(2023)}]{NAP26141}
{National Academies of Sciences Engineering and Medicine}. 2023, Pathways to Discovery in Astronomy and Astrophysics for the 2020s (Washington, DC: The National Academies Press), \dodoi{10.17226/26141}

\bibitem[{Pidhorodetska {et~al.}(2020)Pidhorodetska, Fauchez, Villanueva, Domagal-Goldman, \& Kopparapu}]{Pidhorodetska2020-ip}
Pidhorodetska, D., Fauchez, T., Villanueva, G., Domagal-Goldman, S., \& Kopparapu, R.~K. 2020, Astrophys. J. Lett., 898, L33, \dodoi{10.3847/2041-8213/aba4a1}

\bibitem[{Pilcher(2003)}]{pilcher2003}
Pilcher, C.~B. 2003, Astrobiology, 3, 471, \dodoi{10.1089/15311070332261058}

\bibitem[{Quanz {et~al.}(2022)Quanz, Ottiger, Fontanet, Kammerer, Menti, Dannert, Gheorghe, Absil, Airapetian, Alei, Allart, Angerhausen, Blumenthal, Buchhave, Cabrera, Carri{\'o}n-Gonz{\'a}lez, Chauvin, Danchi, Dandumont, Defr{\'e}re, Dorn, Ehrenreich, Ertel, Fridlund, Garc{\'\i}a~Mu{\~n}oz, Gasc{\'o}n, Girard, Glauser, Grenfell, Guidi, Hagelberg, Helled, Ireland, Janson, Kopparapu, Korth, Kozakis, Kraus, L{\'e}ger, Leedj{\"a}rv, Lichtenberg, Lillo-Box, Linz, Liseau, Loicq, Mahendra, Malbet, Mathew, Mennesson, Meyer, Mishra, Molaverdikhani, Noack, Oza, Pall{\'e}, Parviainen, Quirrenbach, Rauer, Ribas, Rice, Romagnolo, Rugheimer, Schwieterman, Serabyn, Sharma, Stassun, Szul{\'a}gyi, Wang, Wunderlich, \& Wyatt}]{Quanz2022-oo}
Quanz, S.~P., Ottiger, M., Fontanet, E., {et~al.} 2022, Astron. Astrophys. Suppl. Ser., 664, A21, \dodoi{10.1051/0004-6361/202140366}

\bibitem[{Ranjan {et~al.}(2023)Ranjan, Schwieterman, Leung, Harman, \& Hu}]{Ranjan2023-hy}
Ranjan, S., Schwieterman, E.~W., Leung, M., Harman, C.~E., \& Hu, R. 2023, ApJL, 958, L15, \dodoi{10.3847/2041-8213/ad037c}

\bibitem[{Redeker {et~al.}(2000)Redeker, Wang, Low, McMillan, Tyler, \& Cicerone}]{Redeker2000-kj}
Redeker, K.~R., Wang, N., Low, J.~C., {et~al.} 2000, Science, 290, 966, \dodoi{10.1126/science.290.5493.966}

\bibitem[{Rhew \& Abel(2007)}]{Rhew2007-wk}
Rhew, R.~C., \& Abel, T. 2007, Environ. Sci. Technol., 41, 7837, \dodoi{10.1021/es0711011}

\bibitem[{Rigby \& Madhusudhan(2024)}]{Rigby2024-tq}
Rigby, F.~E., \& Madhusudhan, N. 2024, Mon. Not. R. Astron. Soc., 529, 409, \dodoi{10.1093/mnras/stae413}

\bibitem[{Rivera {et~al.}(2005)Rivera, Lissauer, Paul~Butler, Marcy, Vogt, Fischer, Brown, Laughlin, \& Henry}]{Rivera2005}
Rivera, E.~J., Lissauer, J.~J., Paul~Butler, R., {et~al.} 2005, ApJ, 634, 625, \dodoi{10.1086/491669}

\bibitem[{Rugheimer {et~al.}(2013)Rugheimer, Kaltenegger, Zsom, Segura, \& Sasselov}]{Rugheimer2013-je}
Rugheimer, S., Kaltenegger, L., Zsom, A., Segura, A., \& Sasselov, D. 2013, Astrobiology, 13, 251, \dodoi{10.1089/ast.2012.0888}

\bibitem[{Sanz-Novo {et~al.}(2025)Sanz-Novo, Rivilla, Endres, Lattanzi, Jim'enez-Serra, Colzi, Zeng, Meg'ias, L'opez-Gallifa, Mart'inez-Henares, Andr'es, Tercero, de~Vicente, Mart'in, Requena-Torres, Caselli, \& Mart'in-Pintado}]{Sanz-Novo2025-ah}
Sanz-Novo, M., Rivilla, V., Endres, C.~P., {et~al.} 2025, arXiv e-prints, arXiv

\bibitem[{Schwieterman \& Leung(2024)}]{SchwietermanLeung2024}
Schwieterman, E.~W., \& Leung, M. 2024, Reviews in Mineralogy and Geochemistry, 90, 465, \dodoi{10.2138/rmg.2024.90.13}

\bibitem[{Segura {et~al.}(2005)Segura, Kasting, Meadows, Cohen, Scalo, Crisp, Butler, \& Giovanna}]{Segura2005-jv}
Segura, A., Kasting, J.~F., Meadows, V., {et~al.} 2005, Astrobiology, 5, 706, \dodoi{10.1089/ast.2007.0153}

\bibitem[{Sharma {et~al.}(2017)Sharma, Stello, Buder, Kos, Bland-Hawthorn, Asplund, Duong, Lin, Lind, Ness, Huber, Zwitter, Traven, Hon, Kafle, Khanna, Saddon, Anguiano, Casey, Freeman, Martell, De~Silva, Simpson, Wittenmyer, \& Zucker}]{Sharma2017-mm}
Sharma, S., Stello, D., Buder, S., {et~al.} 2017, Mon. Not. R. Astron. Soc., 473, 2004, \dodoi{10.1093/mnras/stx2582}

\bibitem[{Shibazaki {et~al.}(2016)Shibazaki, Ambiru, Kurihara, Tamegai, \& Hashimoto}]{Shibazaki2016-le}
Shibazaki, A., Ambiru, K., Kurihara, M., Tamegai, H., \& Hashimoto, S. 2016, Mar. Chem., 181, 44, \dodoi{10.1016/j.marchem.2016.03.004}

\bibitem[{Shorttle {et~al.}(2024)Shorttle, Jordan, Nicholls, Lichtenberg, \& Bower}]{Shorttle2024-vf}
Shorttle, O., Jordan, S., Nicholls, H., Lichtenberg, T., \& Bower, D.~J. 2024, ApJL, 962, L8, \dodoi{10.3847/2041-8213/ad206e}

\bibitem[{Suissa {et~al.}(2020)Suissa, Mandell, Wolf, Villanueva, Fauchez, \& Kopparapu}]{Suissa2020-sc}
Suissa, G., Mandell, A.~M., Wolf, E.~T., {et~al.} 2020, ApJ, 891, 58, \dodoi{10.3847/1538-4357/ab72f9}

\bibitem[{Tsai {et~al.}(2021{\natexlab{a}})Tsai, Innes, Lichtenberg, Taylor, Malik, Chubb, \& Pierrehumbert}]{Tsai2021-ag}
Tsai, S.-M., Innes, H., Lichtenberg, T., {et~al.} 2021{\natexlab{a}}, ApJL, 922, L27, \dodoi{10.3847/2041-8213/ac399a}

\bibitem[{Tsai {et~al.}(2024)Tsai, Innes, Wogan, \& Schwieterman}]{Tsai2024-gw}
Tsai, S.-M., Innes, H., Wogan, N.~F., \& Schwieterman, E.~W. 2024, ApJL, 966, L24, \dodoi{10.3847/2041-8213/ad3801}

\bibitem[{Tsai {et~al.}(2017)Tsai, Lyons, Grosheintz, Rimmer, Kitzmann, \& Heng}]{Tsai2017-nw}
Tsai, S.-M., Lyons, J.~R., Grosheintz, L., {et~al.} 2017, ApJS, 228, 20, \dodoi{10.3847/1538-4365/228/2/20}

\bibitem[{Tsai {et~al.}(2021{\natexlab{b}})Tsai, Malik, Kitzmann, Lyons, Fateev, Lee, \& Heng}]{Tsai2021-sk}
Tsai, S.-M., Malik, M., Kitzmann, D., {et~al.} 2021{\natexlab{b}}, ApJ, 923, 264, \dodoi{10.3847/1538-4357/ac29bc}

\bibitem[{{Tsai} {et~al.}(2024){Tsai}, {Parmentier}, {Mendon{\c{c}}a}, {Tan}, {Deitrick}, {Hammond}, {Savel}, {Zhang}, {Pierrehumbert}, \& {Schwieterman}}]{Tsai2024-2D}
{Tsai}, S.-M., {Parmentier}, V., {Mendon{\c{c}}a}, J.~M., {et~al.} 2024, \apj, 963, 41, \dodoi{10.3847/1538-4357/ad1600}

\bibitem[{Valencia {et~al.}(2007)Valencia, Sasselov, \& O'Connell}]{Valencia2007-sg}
Valencia, D., Sasselov, D.~D., \& O'Connell, R.~J. 2007, ApJ, 656, 545, \dodoi{10.1086/509800}

\bibitem[{van Pée \& Unversucht(2003)}]{Van_Pee2003-zn}
van Pée, K.~H., \& Unversucht, S. 2003, Chemosphere, 52, 299, \dodoi{10.1016/S0045-6535(03)00204-2}

\bibitem[{Van~Rossum \& Drake(2009)}]{python}
Van~Rossum, G., \& Drake, F.~L. 2009, Python 3 Reference Manual (Scotts Valley, CA: CreateSpace)

\bibitem[{Villanueva {et~al.}(2022)Villanueva, Liuzzi, Faggi, Protopapa, Kofman, Fauchez, Stone, \& Mandell}]{Villanueva2022-lo}
Villanueva, G.~L., Liuzzi, G., Faggi, S., {et~al.} 2022, Fundamentals of the Planetary Spectrum Generator (ui.adsabs.harvard.edu)

\bibitem[{Villanueva {et~al.}(2018)Villanueva, Smith, Protopapa, Faggi, \& Mandell}]{Villanueva2018-jt}
Villanueva, G.~L., Smith, M.~D., Protopapa, S., Faggi, S., \& Mandell, A.~M. 2018, J. Quant. Spectrosc. Radiat. Transf., 217, 86, \dodoi{10.1016/j.jqsrt.2018.05.023}

\bibitem[{Villanueva {et~al.}(2023)Villanueva, Hammel, Milam, Faggi, Kofman, Roth, Hand, Paganini, Stansberry, Spencer, Protopapa, Strazzulla, Cruz-Mermy, Glein, Cartwright, \& Liuzzi}]{Villanueva2023-hb}
Villanueva, G.~L., Hammel, H.~B., Milam, S.~N., {et~al.} 2023, Science, 381, 1305, \dodoi{10.1126/science.adg4270}

\bibitem[{von Bloh {et~al.}(2009)von Bloh, Cuntz, Schr{\"o}der, Bounama, \& Franck}]{Von_Bloh2009-jh}
von Bloh, W., Cuntz, M., Schr{\"o}der, K.-P., Bounama, C., \& Franck, S. 2009, Astrobiology, 9, 593, \dodoi{10.1089/ast.2008.0285}

\bibitem[{Wagg \& Broekgaarden(2024)}]{software-citation-station-zenodo}
Wagg, T., \& Broekgaarden, F. 2024, The Software Citation Station,  Zenodo, \dodoi{10.5281/zenodo.11292917}

\bibitem[{{Wagg} \& {Broekgaarden}(2024)}]{software-citation-station-paper}
{Wagg}, T., \& {Broekgaarden}, F.~S. 2024, arXiv e-prints, arXiv:2406.04405.
\newblock \doarXiv{2406.04405}

\bibitem[{Wogan {et~al.}(2024)Wogan, Batalha, Zahnle, Krissansen-Totton, Tsai, \& Hu}]{Wogan2024-af}
Wogan, N.~F., Batalha, N.~E., Zahnle, K.~J., {et~al.} 2024, ApJL, 963, L7, \dodoi{10.3847/2041-8213/ad2616}

\bibitem[{Xiao {et~al.}(2010)Xiao, Prinn, Fraser, Simmonds, Weiss, O'Doherty, Miller, Salameh, Harth, Krummel, Porter, M{\"u}hle, Greally, Cunnold, Wang, Montzka, Elkins, Dutton, Thompson, Butler, Hall, Reimann, Vollmer, Stordal, Lunder, Maione, Arduini, \& Yokouchi}]{Xiao2010-kd}
Xiao, X., Prinn, R.~G., Fraser, P.~J., {et~al.} 2010, Atmos. Chem. Phys., 10, 5515, \dodoi{10.5194/acp-10-5515-2010}

\bibitem[{Yang \& Hu(2024)}]{Yang2024-pv}
Yang, J., \& Hu, R. 2024.
\newblock \doarXiv{2406.01955}

\bibitem[{Yang {et~al.}(2005)Yang, Cox, Warwick, Pyle, Carver, O'Connor, \& Savage}]{Yang2005-pt}
Yang, X., Cox, R.~A., Warwick, N.~J., {et~al.} 2005, J. Geophys. Res. D: Atmos., 110, 1, \dodoi{10.1029/2005JD006244}

\bibitem[{Yu {et~al.}(2021)Yu, Moses, Fortney, \& Zhang}]{Yu2021-qb}
Yu, X., Moses, J.~I., Fortney, J.~J., \& Zhang, X. 2021, Astrophys. J., 914, 38, \dodoi{10.3847/1538-4357/abfdc7}

\bibitem[{Ziska {et~al.}(2013)Ziska, Quack, Abrahamsson, Archer, {E. Atlas}, Bell, Butler, Carpenter, Jones, Harris, Hepach, Heumann, Hughes, Kuss, Kr{\"u}ger, Liss, Moore, Orlikowska, Raimund, Reeves, Reifenh{\"a}user, Robinson, Schall, Tanhua, Tegtmeier, Turner, Wang, Wallace, Williams, Yamamoto, Yvon-Lewis, \& Yokouchi}]{Ziska2013-tz}
Ziska, F., Quack, B., Abrahamsson, K., {et~al.} 2013, Atmos. Chem. Phys., 13, 8915, \dodoi{10.5194/acp-13-8915-2013}

\end{thebibliography}
\bibliographystyle{aasjournal}



\end{document}